\documentclass[10pt,twocolumn,showpacs,amsmath,amssymb,osajnl,floatfix]{revtex4-1}
\usepackage{epsfig}
\usepackage{amsmath}
\usepackage{amssymb}
\usepackage{bm}
\usepackage{graphicx}
\usepackage{color}
\usepackage{graphicx,rotating}
\usepackage{txfonts}
\usepackage{microtype}
\usepackage{ulem}

\begin{document}
\author{Jian-Xing Li}\email{Jian-Xing.Li@mpi-hd.mpg.de}
\author{Yue-Yue Chen}
\author{Karen Z. Hatsagortsyan}\email{k.hatsagortsyan@mpi-hd.mpg.de}
\author{Christoph H. Keitel}
\affiliation{Max-Planck-Institut f\"{u}r Kernphysik, Saupfercheckweg 1,
69117 Heidelberg, Germany}

\bibliographystyle{apsrev4-1}

\title{Single-shot carrier-envelope phase determination of long superintense laser pulses}

\date{\today}

\begin{abstract}

The impact of the carrier-envelope phase (CEP) of an intense  multi-cycle laser pulse on the radiation of an electron beam during nonlinear Compton scattering is investigated. An interaction regime of the electron beam counterpropagating to the laser pulse is employed, when pronounced high-energy x-ray double peaks emerge at different angles near the backward direction  relative  to the initial electron motion. This is achieved in the relativistic interaction domain, with the additional requirements that
the electron energy is much lower than that necessary for the electron reflection condition at the laser peak, and the stochasticity effects in the photon emission are  weak. The asymmetry parameter of the double peaks in the angular radiation distribution is shown to serve as a sensitive and uniform measure for the CEP of the laser pulse. The method demonstrates unprecedented sensitivity to subtle CEP-effects up to 10-cycle laser pulses and can be applied for the characterization of extremely strong laser pulses in present and near future laser facilities.


\end{abstract}

\maketitle

Superintense laser technique has been developing rapidly in recent years. Petawatt (PW) laser systems are developed across the globe, and 10 PW short-pulse lasers are anticipated in near future, e.g., in the Central Laser Facility in UK (Vulcan) \cite{Vulcan10},  in the Extreme Light Infrastructure (ELI) Nuclear Physics (ELI-NP) in Romania \cite{ELI-NP}, and in the ELI-Beamlines in the Czech Republic \cite{ELI-Beamlines}. Thus,  the present record of a laser intensity of $I\sim 10^{22}$ W/cm$^2$ \cite{Yanovsky_2008} soon will be widely available, and laser projects for intensities as high as $10^{23}-10^{25}$ W/cm$^2$ are under construction, e.g., ELI and Exawatt Center  
for Extreme Light Studies (XCELS) \cite{ELI,XCELS}, opening bright prospect for investigation of new regimes of  laser-matter interaction \cite{Marklund_2006,Mourou_2006,Esarey_2009,RMP_2012,Mourou_2014}. 

Extremely intense lasers require new techniques for characterization of laser-pulse parameters: intensity, focal radius, pulse shape, chirp, and carrier-envelope phase (CEP). The CEP is an important parameter in the strong field physics and nonlinear QED.
Thus, the CEP has a significant impact on the electron spin \cite{Meuren_2011}, the angular distribution, asymmetry, and cross-section of nonlinear  Compton scattering \cite{Boca_2009,Mackenroth_2010,Seipt_2013},  and of the electron-positron pair production process via different mechanisms \cite{Hebenstreit_2009,Krajewska_2012,Nuriman_2013,Titov_2016,Jansen_2016,Meuren_2016}. 
In general, the CEP provides a useful handle to control the physical properties of the laser-matter interaction.
When the laser intensity is below the relativistic threshold ($I\sim 10^{18}$ W/cm$^2$), 
the CEP can be determined via employing asymmetry of above-threshold ionization
\cite{paulus_2001,Paulus_2003, Wittmann_2009}, 
probing the variation of the field strength with the streaking method \cite{Goulielmakis_2004}, or applying terahertz-emission spectroscopy \cite{Kress_2006}. In the relativistic domain of laser intensities, it was shown that signatures of the CEP of  few-cycle laser pulses can be detected via  nonlinear Compton scattering from either the bandwidth of the angular distribution of the electron radiation \cite{Mackenroth_2010} or the differential cross sections of the Breit-Wheeler pair production process \cite{Titov_2016}. However, the discussed CEP effects have demonstrated  high sensitivity only for ultrashort laser pulses with a duration up to at most two cycles. Currently achievable ultra-intense laser pulses and especially those under construction of intensity $I\sim10^{22}-10^{25}$ W/cm$^2$ and of pulse duration $\sim 20-30$ fs though consist of about $6-10$ cycles. In particular, this holds for the $\sim30$ fs pulses of Vulcan \cite{Vulcan10}, the $\sim25$ fs pulses of ELI-NP \cite{ELI-NP} as well as the ones of length $\sim15-20$ fs in ELI-Beamlines \cite{ELI-Beamlines}, of $\sim30$ fs in Ref.~\cite{Yanovsky_2008} and of $\sim25$ fs in XCELS \cite{XCELS}. Thus, there is apparent need for identifying novel methods to precisely characterize the CEP for multi-cycle ultra-intense lasers.

In this letter, we propose a method for measuring CEP of intense multi-cycle laser pulses. The method is based on a sensitive signature of CEP in the subtle features of the angle-resolved radiation spectra of the electron beam via nonlinear Compton scattering. 
A setup is considered when a relativistic electron beam initially counterpropagates to an intense multi-cycle laser pulse. We judiciously choose the regime when the  backward radiation relative to the electron's initial motion is enhanced, forming a broad peak splitting into two parts.  The asymmetry parameter of these  two peaks provides a sensitive measure of CEP. The designated regime is achieved in the relativistic domain, however, with a rather small Lorentz factor $\gamma$ of the electrons, such that the  interaction with the laser field is below the, so-called, reflection condition \cite{DiPiazza_2009}. Moreover,  the stochasticity effects are required to be rather weak, opposite to the case considered in \cite{Jianxing_2017}.

Defining more concretely the parameters of the considered regime: for the below reflection condition $\gamma$ should be much smaller than the invariant laser intensity parameter $\xi$: $\gamma\ll\xi$, where $\xi\equiv eE_0/(m\omega_0)$, $E_0$ and $\omega_0$ are the amplitude and frequency of the laser field, respectively, and $-e$ and $m$  the electron charge and mass, respectively. Planck units $\hbar=c=1$ are used throughout.  For weak stochasticity effects, $\chi\lesssim 0.1$ is required \cite{Shen_1972,Duclous_2011,Neitz_2013}, where $\chi\equiv |e|\sqrt{(F_{\mu\nu}p^{\nu})^2}/m^3$ is the invariant quantum parameter \cite{Reiss_1962,Ritus_1985}, $F_{\mu\nu}$ the field tensor, and $p^{\nu}=(\varepsilon,\textbf{p})$  the incoming electron 4-momentum.

\begin{figure}
\includegraphics[width=\linewidth]{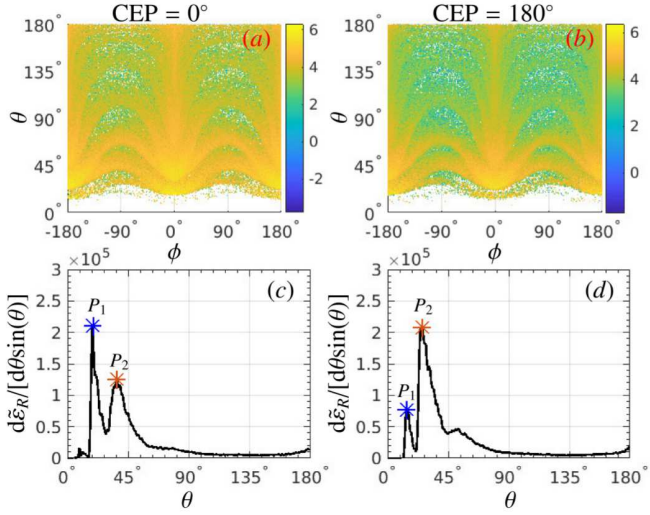}
\caption{ (Color online) (a) and (b): angle-resolved radiation energy $\varepsilon_R$ in units of the electron rest energy $m$ vs the emission polar angle $\theta$ and the azimuthal angle $\phi$  in a 6-cycle focused laser pulse; color coded is log$_{10}$[d$\varepsilon_R/$d$\Omega$] rad$^{-2}$ with the emission solid angle $\Omega$; the CEP $\psi_{\rm CEP}=0^{\circ}$ and $180^{\circ}$, respectively. (c) and (d): the variation of d$\tilde{\varepsilon}_R$/[d$\theta$sin($\theta$)] with respect to $\theta$  taking $\psi_{\rm CEP}=0^{\circ}$ and $180^{\circ}$, respectively; $P_1$ and $P_2$ are corresponding to the two main peaks from left to right. All other parameters are given in the text.   }
\label{fig1}
\end{figure}

In the common setup of nonlinear Compton scattering  \cite{Bula_1996,Panek_2002,Seipt_2011,Vranic_2016}
the condition $\gamma\gg\xi$ for the initial laser and electron parameters is employed,
when the radiation concentrates mainly in the forward direction relative to the initial motion of electrons. In the regime of interaction close to the reflection condition $\gamma\sim\xi/2$, backward emission appears in the angle-resolved radiation spectrum \cite{DiPiazza_2009,Neitz_2013,Mackenroth_2010, Li_2015}. When the parameters controlling radiation reaction and laser focusing are adapted in such a way that the reflection takes place at the peak of the laser pulse in the focal spot, a broad peak arises in the backward radiation \cite{Jianxing_2017}.  
In this paper $\gamma\ll\xi$, when the electron entering the counterpropagating laser beam is reflected before reaching the laser field peak and subsequently accelerated along the laser pulse.
Before the reflection the forward radiation is weak due to the small $\gamma$ and the low laser field. The electron mainly radiates backwards 
after acceleration when it experiences the near peak region of the laser pulse. Multiple bursts of radiation arise in the angle-resolved spectra in the backward direction, which correspond to the laser-cycle structure.  While in the quantum 
regime of  $\chi\sim 1$ the multiple bursts coalesce into a single backward peak due to stochasticity effects of the photon emission \cite{Jianxing_2017},  here we use  $\chi\ll 1$ limit when at least two radiation peaks are well exhibited in the angle-resolved backward spectrum, see Fig.~\ref{fig1}. These peaks sensitively probe the structure of the laser pulse. Consequently, the asymmetry of the peaks is significant even in the case of multi-cycle laser pulses. The asymmetry parameter of the peaks 
monotonously varies with respect to CEP, allowing to measure CEP of multi-cycle laser pulses.
 
The electron radiation is simulated using the QED Monte-Carlo approach, applicable in superstrong laser fields $\xi\gg 1$ \cite{Elkina_2011,Ridgers_2014,Green_2015,Suppl_material}. The photon emission probability in this limit is determined by the local value of the parameter $\chi$ \cite{Baier_b_1994}.  Between photon emissions electrons are propagated via classical equations of motion.    
In our simulations, even though the 
parameter $\chi$ is small, 
the discrete and probabilistic character of photon emission is accounted for.
We employ a linearly polarized focused laser pulse with a Gaussian temporal profile and the CEP $\psi_{\rm CEP}$, 
which propagates along $+z$-direction and is polarized in $x$-direction (for details on the configuration and pulse structure see \cite{Suppl_material}). 
 The spatial distribution of the electromagnetic fields takes into account up to the $\epsilon^3$-order of the nonparaxial solution \cite{Salamin_2002, Suppl_material}, where the focusing parameter $\epsilon=w_0/z_r$, 
$w_0$ is the laser focal radius, $z_r = \pi w_0^2/\lambda_0$  the Rayleigh length, and $\lambda_0$  the laser wavelength. 
In the laser-electron interaction, the quantum invariant parameter $\chi= \gamma(\omega_0/m)\xi (1-\beta \cos\theta)\approx10^{-6}\gamma\xi\lesssim 10^{-1}$, where $\beta$ is the Lorentz $\beta$-parameter, and $\theta$ the angle between the laser wave vector and electron momentum. The required conditions  $\xi\gg \gamma\gg 1$ and $\chi\ll 1$ are fulfilled at $\xi\sim 10^2-10^3$ ($I \sim 10^{22}$-$10^{25}$ W/cm$^2$).
We consider here the currently realistic laser intensities of $I\sim 10^{22}$-$10^{23}$ W/cm$^2$.

 \begin{figure}[b]
\includegraphics[width=8cm]{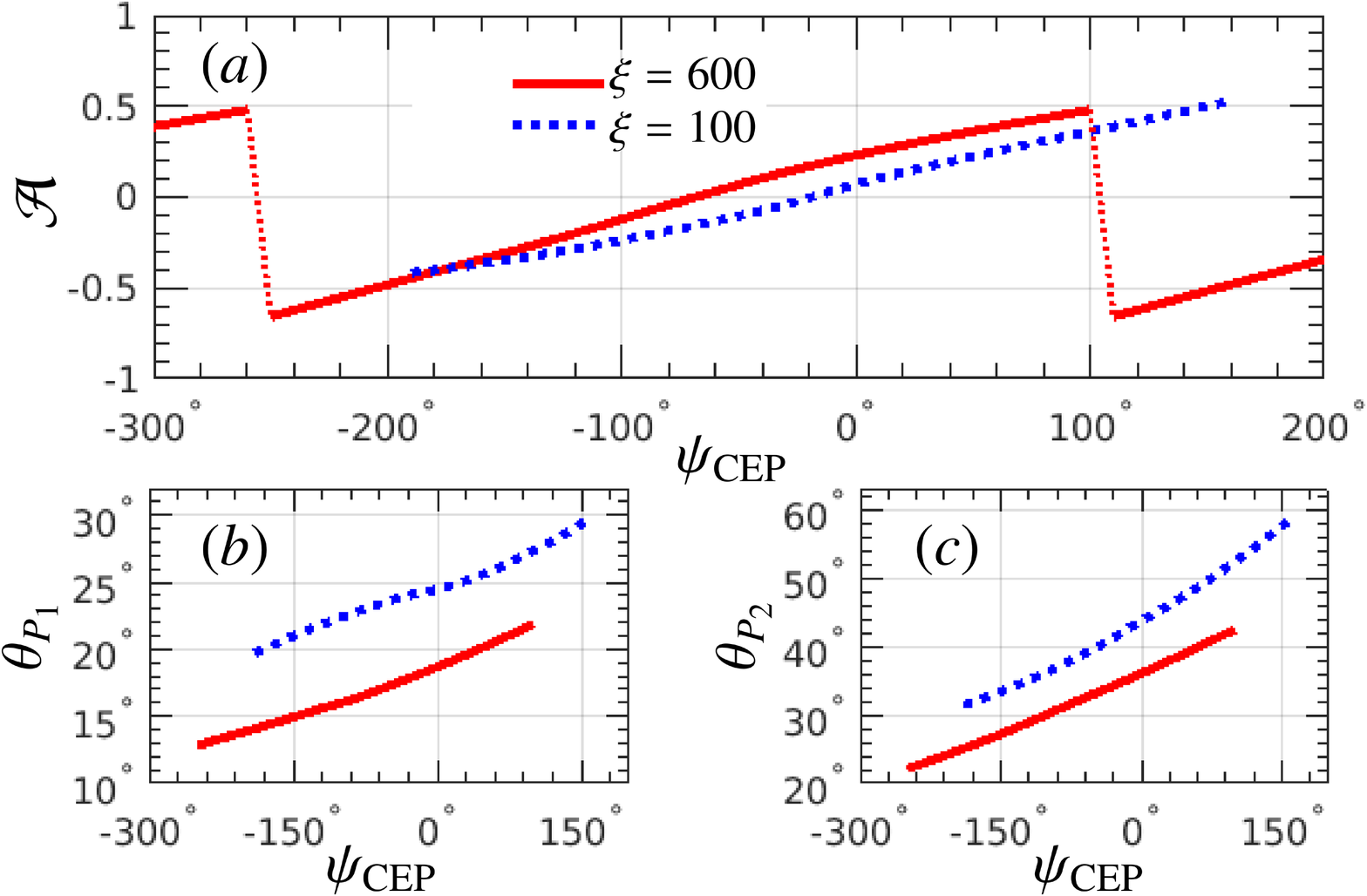}
\caption{  (a) The asymmetry parameter ${\cal A}$ of the backward radiation peaks $P_1$ and $P_2$ vs CEP. The polar angles (b) $\theta_{P_1}$ and (c) $\theta_{P_2}$ vs CEP. The red-solid and blue-dotted curves represent the results of the laser intensities of $\xi=600$ and 100, respectively. The electron kinetic energies are 10 MeV and 3 MeV, respectively.  The periodic variation of ${\cal A}$ for $\xi=600$ is shown in (a), and  are omitted for other cases.  Other parameters are the same as in Fig.~\ref{fig1}.}
\label{fig2}
\end{figure}

The angular distributions of radiation in 6-cycle (FWHM) laser pulses with different CEPs are illustrated in Fig.~\ref{fig1}. The laser and electron beam propagate with a initial polar angle $\theta_L=0^{\circ}$ and $\theta_e=179^{\circ}$, respectively. The  peak intensity of the laser pulse is $I\approx 4.9\times 10^{23}$W/cm$^2$ ($\xi=600$), $\lambda_0$ = 1 $\mu$m, and $w_0$ = 2 $\mu$m. The electron beam parameters are typical for the laser-plasma acceleration setup \cite{Esarey_2009}. The electron beam  radius is $w_e=\lambda_0 $ and length $L_e=6 \lambda_0$, and the total electron number is $N_e=1.2\times10^5$ (electron density $n_e\approx 6.37\times 10^{15}$ cm$^{-3}$).  The initial mean kinetic energy of the electron is $\varepsilon_0 =10$ MeV ($\gamma_0\approx 19.6$, the maximum value of $\chi$ during interaction $\chi_{\rm max}\approx 0.037$), and the energy and angular spread  are $\Delta \varepsilon/\varepsilon_0=\Delta \theta=0.02$. 
In the considered linearly polarized laser pulse, the azimuthal angle $\phi=0^{\circ}$ and $\pm 180^{\circ}$ correspond to the positive and negative directions of the polarization, respectively.  The radiation around $0^{\circ}$ and $180^{\circ}$ are not symmetric due to asymmetry of the laser pulse.

The relativistic electrons penetrate into the laser field, however, the forward radiation is rather weak since the initial $\chi\sim 10^{-2}$ is very small.
As the electrons are reflected and accelerated by the intense laser field, the radiation which is in the backward direction relative to the electron initial motion, is enhanced. This is because the parameters $\gamma$, $\xi$, and instantaneous emitted photon energy $\varepsilon_{\gamma}\sim\gamma\chi$ are increased.
During the reflection, the emission polar angle $\theta$ varies from $180^{\circ}$ to close to $0^{\circ}$.

 The angle-resolved spectra of the radiation significantly depend on the CEP.  To quantify the CEP effect, we focus on the strongest radiation domain along the polarization plane in the region of $-15^{\circ}\leq \phi\leq +15^{\circ}$, analyzing the radiation energy   d$\tilde{\varepsilon}_R$/[d$\theta$sin($\theta$)]= $\int_{-15^{\circ}}^{+15^{\circ}}$ d$\phi$  d$\varepsilon_R$/d$\Omega$, as shown in Figs.~\ref{fig1}(c) and \ref{fig1}(d). The two main peaks of the radiation are marked as $P_1$ and $P_2$. 
 The relative height of the peaks and the corresponding polar angles are different at $\psi_{\rm CEP}=0^{\circ}$ and $\psi_{\rm CEP}=180^{\circ}$.
We define the asymmetry parameter of the peaks
\begin{eqnarray}
{\cal A}=\frac{M_{P_1}-M_{P_2}}{M_{P_1}+M_{P_2}},
\end{eqnarray}
with the height of the peaks $M_{P_{1,2}}=d\tilde{\varepsilon}_R$/[d$\theta$sin($\theta)]|_{\theta=\theta_{P_{1,2}}}$, and the corresponding polar angles  $\theta_{P_1}$ and $\theta_{P_2}$, respectively.
 
\begin{figure}
\includegraphics[width=8cm]{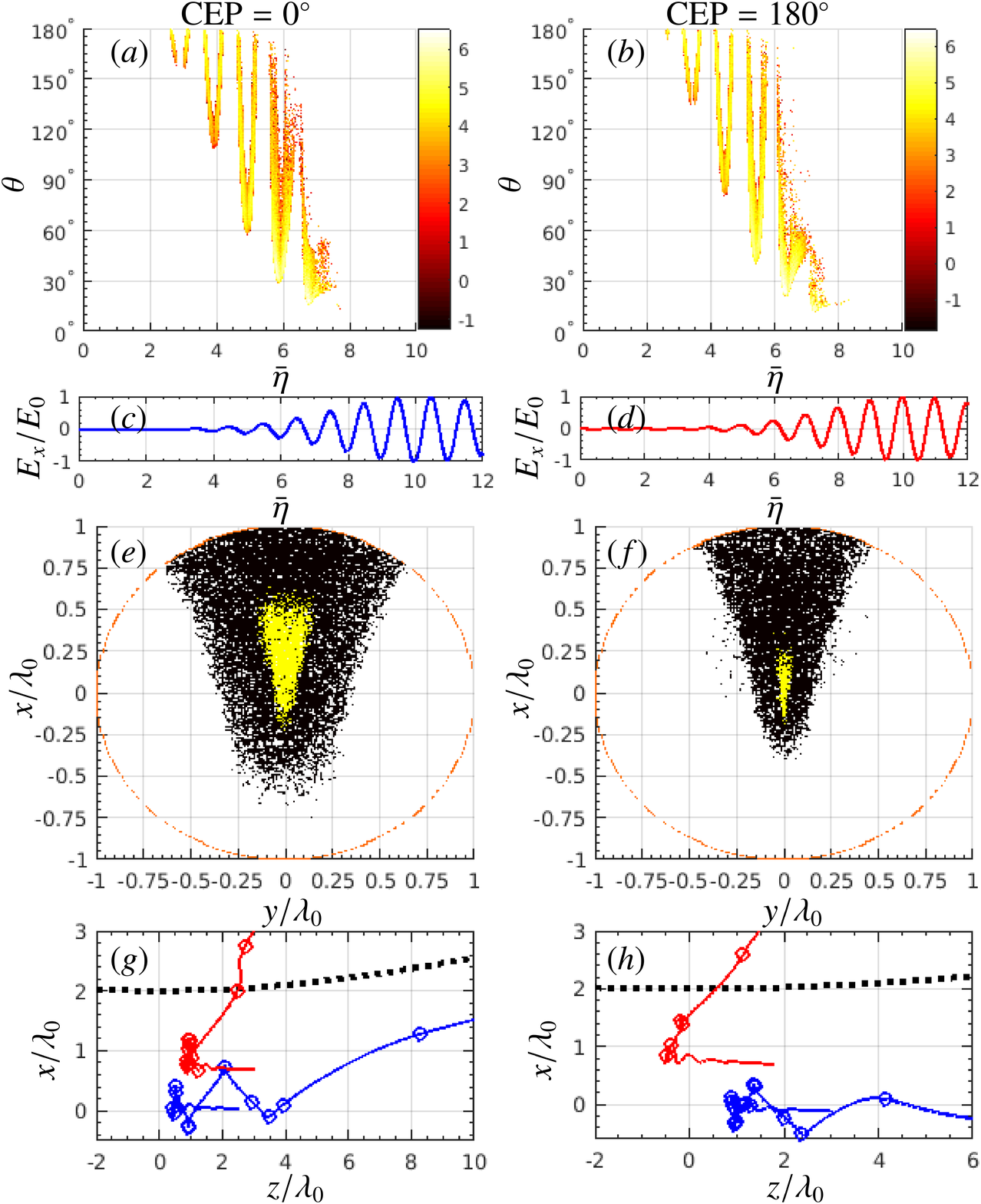}
\caption{(Color online) Emergence of the radiation peaks $P_1,P_2$ (see Fig.~\ref{fig1}): Left column is for $\psi_{\rm CEP}=0^{\circ}$, and right column  for $\psi_{\rm CEP}=180^{\circ}$. (a) and (b): radiation intensity integrated over the azimuthal angle of $-15^{\circ}\leq \phi\leq +15^{\circ}$, $\log_{10}\{d^2\tilde{\varepsilon}_R/[d\bar{\eta}d\theta \sin(\theta)]\}$, vs emission phase $\bar{\eta}$, , with  $\bar{\eta}= (\omega_0 t -k_0z)/2\pi$. (c) and (d): the transverse component of electric fields $E_x$ at the focus scaled by the laser amplitude $E_0$ vs $\bar{\eta}$. (e) and (f): the electron initial spatial distribution in the cross section of the electron beam, which contributes to the spectral peak $P_1$ (yellow), and to the spectral peak $P_2$ (black) (the latter includes also the yellow region). 
The red circles show the boundary of the electron beam. (g) and (h):  example trajectories of the electrons initially in the yellow region (blue curves), and in the black region (outside of the yellow part, red curves), respectively.  The photon emission is indicated by circles, and the dashed line shows the laser beam radius $w_z=w_0\sqrt{1+\left(z/z_r\right)^2}$. Other parameters are the same as in Fig.~\ref{fig1}. }
\label{fig3}
\end{figure}

\begin{figure}[]
\includegraphics[width=8cm]{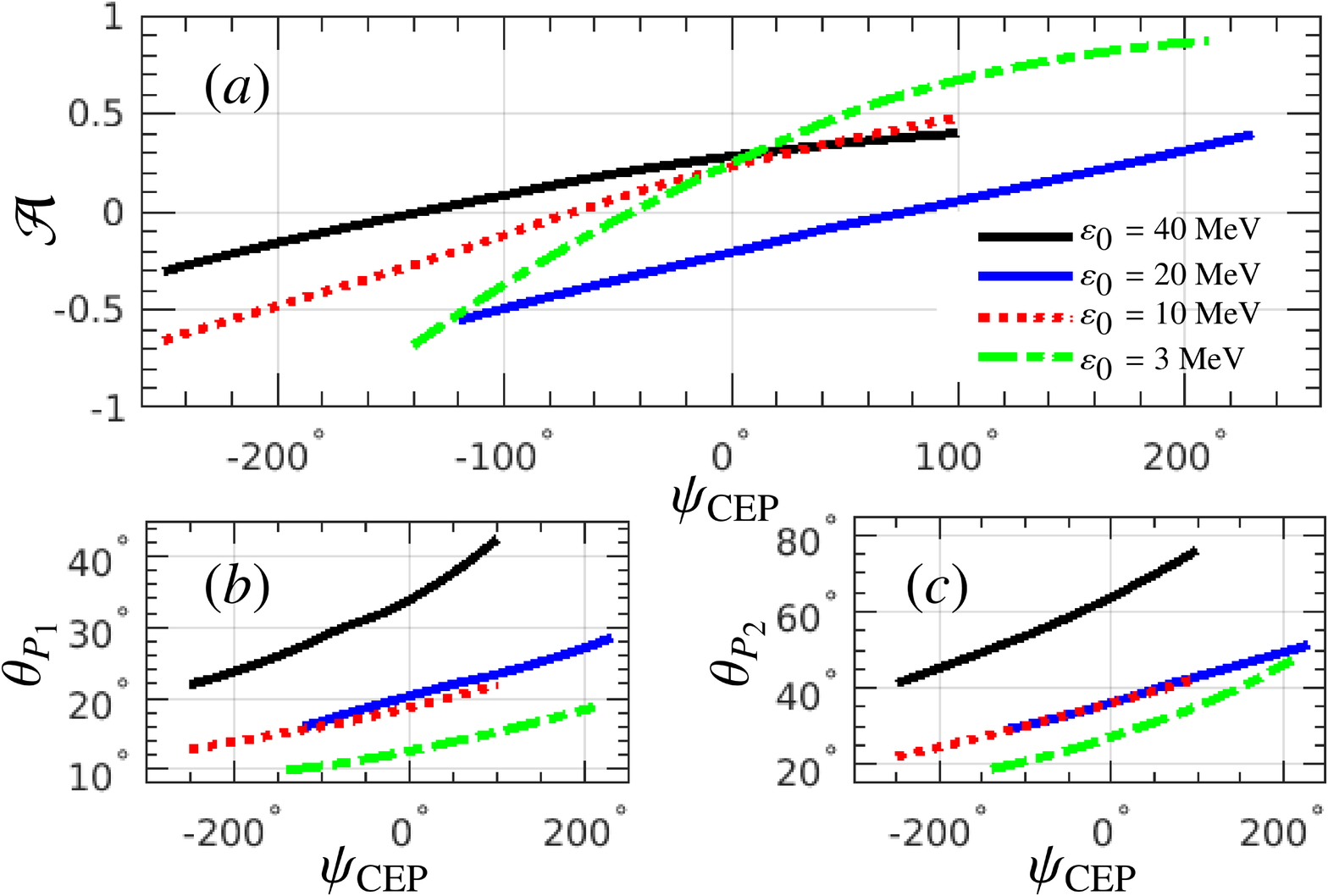}
\caption{(Color online)  (a) The asymmetry parameter ${\cal A}$  vs the CEP phase.   The polar angles (b) $\theta_{P_1}$ and (c) $\theta_{P_2}$ vs the CEP phase. The black-solid, blue-solid, red-dotted and green-dash-dotted curves correspond to $\varepsilon_0=$ 40, 20, 10 and 3 MeV, respectively. Other parameters are the same as in Fig.~\ref{fig1}. }
\label{fig4}
\end{figure}

\begin{figure}
\includegraphics[width=8cm]{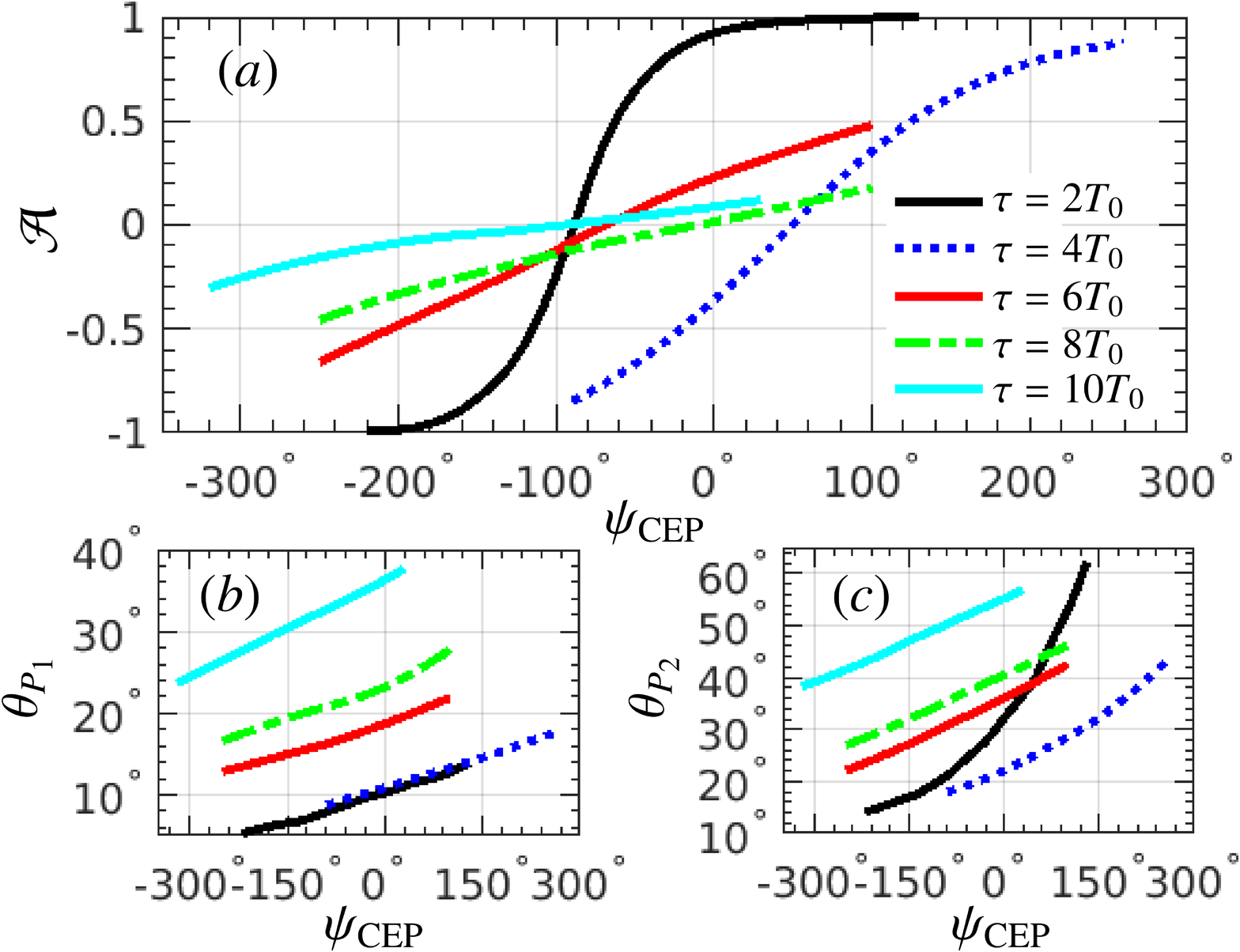}
\caption{(Color online) The variations of (a) ${\cal A}$ between $P_1$ and $P_2$ and the polar angles (b) $\theta_{P_1}$ and (c) $\theta_{P_2}$ with respect to the CEP. The black-solid, blue-dotted, red-solid, green-dash-dotted and cyan-solid curves represent the cases of $\tau$ = 2, 4, 6, 8 and 10 $T_0$, respectively.  Other parameters are the same as in Fig.~\ref{fig1}. }
\label{fig5}
\end{figure}

We proceed the analysis of dependencies of ${\cal A}$, $\theta_{P_1}$ and $\theta_{P_2}$ on CEP with a CEP interval of $10^{^{\circ}}$, as shown in Fig.~\ref{fig2}. And, the results of two laser intensities of $I\sim 10^{22}$ W/cm$^2$ ($\xi=100$, blue-dash curves) and $I\sim10^{23}$ W/cm$^2$ ($\xi=600$, red-solid curves) are compared.
${\cal A}$, $\theta_{P_1}$ and $\theta_{P_2}$ all monotonously increase with $\psi_{\rm CEP}$, which can be used to  characterize CEP of the laser pulse. In particular, as shown in Fig.~\ref{fig2}(a), the asymmetry parameter ${\cal A}$ varies in a large range from approximately $-0.5$ to $0.5$  for both intensities. The emission angles of the peaks also can be used as a CEP indicator. As $\theta_{P_2}$ varies with CEP in a larger range than $\theta_{P_1}$, see Fig.~\ref{fig2}(b) and (c)  [$\theta_{P_1}$ grows approximately by 9.07$^{\circ}$, from 12.81$^{\circ}$ to 21.88$^{\circ}$, for $\xi=600$, and by 10.1$^{\circ}$, from 19.69$^{\circ}$ to 29.79$^{\circ}$, for $\xi=100$;  $\theta_{P_2}$ grows by 20.4$^{\circ}$, from 22.07$^{\circ}$ to 42.47$^{\circ}$, for $\xi=600$, and by 27.22$^{\circ}$, from 31.4$^{\circ}$ to 58.62$^{\circ}$, for $\xi=100$], 
the determination of CEP via $\theta_{P_2}$ is preferable.  

Note that the CEP signatures are not affected much from decreasing the value of the $\chi$-parameter (for our calculations in the $\xi=100$ case the parameter $\chi_{\rm max}\approx0.004$ is much smaller than $\chi_{\rm max}\approx0.037$ for the $\xi=600$ case), although the total radiation intensity is decreased.

We analyze the emergence of radiation peaks and their relative heights in
Fig.~\ref{fig3}.  Figures~\ref{fig3}(a) and \ref{fig3}(b) show the radiation intensity resolved in laser cycles for $\psi_{\rm CEP}=0^{\circ}$ and  $180^{\circ}$, respectively.  In each laser cycle the strongest radiation arises near the peaks of the cycles at a certain emission angle. Between adjacent radiation peaks, there is a gap in the emission polar angle corresponding to the weak field part of the laser cycle. Integrating the radiation intensities in Figs.~\ref{fig3}(a) and \ref{fig3}(b) by the emission laser phase $\bar{\eta}$ generates the peak structures of radiation in Figs.~\ref{fig1}(c) and \ref{fig1}(d). 
As the initial energies of the electrons $\varepsilon_0\ll\xi/2$, the electrons are easily reflected ($\theta \approx 90^{\circ}$) and accelerated by the laser fields before the laser peak, $\bar{\eta}\approx 5$ in Fig.~\ref{fig3}(c), and $\bar{\eta}\approx 4.2$ in Fig.~\ref{fig3}(d). 
Short after the reflection, the pondermotive force due to the transverse profile of the focused laser field pushes the electrons transversely  out of the laser pulses (see the trajectories of the sample electrons in Figs.~\ref{fig3}(g) and \ref{fig3}(h)). The farther the electron is away from the beam center, the faster it is expelled from the beam, and this at a larger angle. In fact, as illustrated in Figs.~\ref{fig3}(e) and \ref{fig3}(f), the first peak $P_1$ at small angles (Fig.~\ref{fig1}) is exclusively formed by the radiation of the electrons initially located in the center of the beam (the yellow area of  the beam), and the second peak $P_2$ by the electrons located far from the beam center (the black area). The reason is that the ponderomotive force, being proportional to the gradient of the transverse profile of the laser beam, is larger for electrons at the wings of the beam than in the center of the beam.  Note, however, that during  oscillation electrons from the yellow region radiate in other polar angles as well. Moreover, we find that  the longitudinal position of the electron in the  beam does not affect significantly the generation of radiation bursts.

We proceed to discuss the impacts of the laser and electron parameters on the CEP signatures. The role of the electron's initial kinetic energy for CEP signatures is analyzed  in Fig. ~\ref{fig4}. As the electron kinetic energy decreases from 40 MeV to 3 MeV, the gradient of ${\cal A}$ increases, although the absolute intensity of the radiation decreases due to the decrease of the $\chi$-parameter. 
In the considered Gaussian laser pulse, around the laser-envelope peak, the laser intensity is inversely proportional to its gradient (see Figs.~\ref{fig3}(c) and \ref{fig3}(d)). A slower electron can be reflected by a lower laser intensity ($\gamma\sim\xi/2$), where the gradient of the envelope is larger. Consequently, the difference of the height of the two radiation peaks corresponding to adjacent laser cycles, i.e., the asymmetry parameter ${\cal A}$, is larger.  This further proves that the condition $\gamma\ll\xi$ is  crucial.  
The appropriate $\gamma$-parameter should be at least one order 
of magnitude smaller than $\xi$. However, the electron's kinetic energy should not be too small, because it would be reflected immediately by the laser prepulse without reaching the laser pulse region with high gradient of the envelope.  The gradients of $\theta_{P_1}$ and $\theta_{P_2}$ do not vary much with the electron energy, and the observation of the CEP signatures with $\theta_{P_2}$ is confirmed to be more beneficial than with $\theta_{P_1}$.

Let us estimate the resolution of our method for the CEP measurement.  With current achievable angular resolution for x-ray registration in  experiments of $\approx 1$ mrad ($\approx 0.057^{\circ}$) \cite{Rousse_2004,Cipiccia_2011,Graves_2014}, the CEP can be measured via $\theta_{P_2}$ ($\theta_{P_2}$ changes by $\approx 20^\circ$ within $\Delta \psi_{\rm CEP}=360^\circ$ in Fig.~\ref{fig4}) with an accuracy of approximately  $1^\circ$ in the case of a 6-cycle laser pulse. For the CEP retrival with the asymmetry parameter ${\cal A}$, we estimate the feasible resolution of ${\cal A}$ as $\Delta{\cal A}\approx\Delta N_{ph}/N_{ph}\sim 1/\sqrt{N_{ph}}$ with the emitted photon number $N_{ph}$, and $N_{ph}\approx  \xi\alpha N_e$ \cite{Ritus_1985}, with the fine-structure constant $\alpha$, since the photon emission at $P_{1,2}$ mainly happens during a single coherence length. In the considered electron beam with $\xi=100$, $\Delta{\cal A}\approx3.4\times10^{-3}$, and the resolution of CEP detection is approximately 1.2$^{\circ}$. With higher electron number $N_e \sim 10^6$, the CEP resolution improves to $\sim 0.4^{\circ}$.

The advantage of the present method of the CEP detection with respect to other methods \cite{Mackenroth_2010,Seipt_2013} is that it is sensitive to CEP effects of relatively long laser pulses. The CEP signatures in dependence of the laser-pulse length are discussed in Fig.~\ref{fig5}. As the laser-pulse duration increases from 2 cycles to 10 cycles, the gradient of the laser-field amplitude along the laser pulse decreases, therefore, the gradient of ${\cal A}$ decreases as well, see Fig.~\ref{fig5}(a). Although the gradient of the asymmetry parameter is large in  the 2-cycle laser pulse in a certain CEP region, in the regions of $\psi_{\rm CEP}\lesssim-200^{\circ}$ and $\psi_{\rm CEP}\gtrsim 100^{\circ}$, ${\cal A}$ saturates  ${\cal A}\approx \mp1$, when 
one of peaks is much smaller than another one. One can deduce from Fig.~\ref{fig5} that in ultrashort pulses less than 2-cycles, $\theta_{P_1}$ becomes more suitable measure of CEP because it maintains uniformity in the full CEP range. The CEP resolutions for the 4-, 6-, 8- and 10-cycle cases with other parameters as in Fig.~\ref{fig1} are approximately 0.36$^{\circ}$, 0.44$^{\circ}$, 0.68$^{\circ}$ and 0.93$^{\circ}$, respectively. Thus, the resolution is oppositely proportional    to the laser-pulse length.

Concluding, we investigate a new method for the determination of the CEP of intense long  laser pulses via analyzing angle-resolved radiation spectra via nonlinear Compton scattering. Two main radiation peaks are generated in backward radiation relative to the electron motion in suitable conditions.
The asymmetry parameter and the corresponding emission polar angles of the two peaks can characterize the CEP of the laser pulse with  a high  resolution of about $1^{\circ}$. The method is robust with the laser and electron beam parameters and are applicable for currently achievable laser sources and those under-construction of relativistic intensities  .   

\bibliography{strong_fields_bibliography}

\begin{thebibliography}{47}%
\makeatletter
\providecommand \@ifxundefined [1]{%
 \@ifx{#1\undefined}
}%
\providecommand \@ifnum [1]{%
 \ifnum #1\expandafter \@firstoftwo
 \else \expandafter \@secondoftwo
 \fi
}%
\providecommand \@ifx [1]{%
 \ifx #1\expandafter \@firstoftwo
 \else \expandafter \@secondoftwo
 \fi
}%
\providecommand \natexlab [1]{#1}%
\providecommand \enquote  [1]{``#1''}%
\providecommand \bibnamefont  [1]{#1}%
\providecommand \bibfnamefont [1]{#1}%
\providecommand \citenamefont [1]{#1}%
\providecommand \href@noop [0]{\@secondoftwo}%
\providecommand \href [0]{\begingroup \@sanitize@url \@href}%
\providecommand \@href[1]{\@@startlink{#1}\@@href}%
\providecommand \@@href[1]{\endgroup#1\@@endlink}%
\providecommand \@sanitize@url [0]{\catcode `\\12\catcode `\$12\catcode
  `\&12\catcode `\#12\catcode `\^12\catcode `\_12\catcode `\%12\relax}%
\providecommand \@@startlink[1]{}%
\providecommand \@@endlink[0]{}%
\providecommand \url  [0]{\begingroup\@sanitize@url \@url }%
\providecommand \@url [1]{\endgroup\@href {#1}{\urlprefix }}%
\providecommand \urlprefix  [0]{URL }%
\providecommand \Eprint [0]{\href }%
\providecommand \doibase [0]{http://dx.doi.org/}%
\providecommand \selectlanguage [0]{\@gobble}%
\providecommand \bibinfo  [0]{\@secondoftwo}%
\providecommand \bibfield  [0]{\@secondoftwo}%
\providecommand \translation [1]{[#1]}%
\providecommand \BibitemOpen [0]{}%
\providecommand \bibitemStop [0]{}%
\providecommand \bibitemNoStop [0]{.\EOS\space}%
\providecommand \EOS [0]{\spacefactor3000\relax}%
\providecommand \BibitemShut  [1]{\csname bibitem#1\endcsname}%
\let\auto@bib@innerbib\@empty
\bibitem [{\citenamefont {{The Vulcan facility}}()}]{Vulcan10}%
  \BibitemOpen
  \bibfield  {author} {\bibinfo {author} {\bibnamefont {{The Vulcan
  facility}}},\ }\href@noop {} {}\bibinfo {howpublished} {\url{https://www.clf.
  stfc.ac.uk/Pages/The-Vulcan-10-Petawatt-Project.aspx.}}\BibitemShut {Stop}%
\bibitem [{\citenamefont {Negoita}\ \emph {et~al.}(2016)\citenamefont
  {Negoita}, \citenamefont {Roth}, \citenamefont {Thirolf},\ and\ \citenamefont
  {et~al.}}]{ELI-NP}%
  \BibitemOpen
  \bibfield  {author} {\bibinfo {author} {\bibfnamefont {F.}~\bibnamefont
  {Negoita}}, \bibinfo {author} {\bibfnamefont {M.}~\bibnamefont {Roth}},
  \bibinfo {author} {\bibfnamefont {P.~G.}\ \bibnamefont {Thirolf}}, \ and\
  \bibinfo {author} {\bibnamefont {et~al.}},\ }\href@noop {} {\bibfield
  {journal} {\bibinfo  {journal} {Rom. Rep. Phys.}\ }\textbf {\bibinfo {volume}
  {68}},\ \bibinfo {pages} {S37} (\bibinfo {year} {2016})}\BibitemShut
  {NoStop}%
\bibitem [{\citenamefont {{ElI-Beamlines}}()}]{ELI-Beamlines}%
  \BibitemOpen
  \bibfield  {author} {\bibinfo {author} {\bibnamefont {{ElI-Beamlines}}},\
  }\href@noop {} {}\bibinfo {howpublished} {\url{https://www.eli-beams.eu/en/
  facility/lasers/}}\BibitemShut {NoStop}%
\bibitem [{\citenamefont {Yanovsky}\ \emph {et~al.}(2008)\citenamefont
  {Yanovsky}, \citenamefont {Chvykov}, \citenamefont {Kalinchenko},
  \citenamefont {Rousseau}, \citenamefont {Planchon}, \citenamefont {Matsuoka},
  \citenamefont {Maksimchuk}, \citenamefont {Nees}, \citenamefont {Cheriaux},
  \citenamefont {Mourou},\ and\ \citenamefont {Krushelnick}}]{Yanovsky_2008}%
  \BibitemOpen
  \bibfield  {author} {\bibinfo {author} {\bibfnamefont {V.}~\bibnamefont
  {Yanovsky}}, \bibinfo {author} {\bibfnamefont {V.}~\bibnamefont {Chvykov}},
  \bibinfo {author} {\bibfnamefont {G.}~\bibnamefont {Kalinchenko}}, \bibinfo
  {author} {\bibfnamefont {P.}~\bibnamefont {Rousseau}}, \bibinfo {author}
  {\bibfnamefont {T.}~\bibnamefont {Planchon}}, \bibinfo {author}
  {\bibfnamefont {T.}~\bibnamefont {Matsuoka}}, \bibinfo {author}
  {\bibfnamefont {A.}~\bibnamefont {Maksimchuk}}, \bibinfo {author}
  {\bibfnamefont {J.}~\bibnamefont {Nees}}, \bibinfo {author} {\bibfnamefont
  {G.}~\bibnamefont {Cheriaux}}, \bibinfo {author} {\bibfnamefont
  {G.}~\bibnamefont {Mourou}}, \ and\ \bibinfo {author} {\bibfnamefont
  {K.}~\bibnamefont {Krushelnick}},\ }\href@noop {} {\bibfield  {journal}
  {\bibinfo  {journal} {Opt. Express}\ }\textbf {\bibinfo {volume} {16}},\
  \bibinfo {pages} {2109} (\bibinfo {year} {2008})}\BibitemShut {NoStop}%
\bibitem [{\citenamefont {{The Extreme Light Infrastructure (ELI)}}()}]{ELI}%
  \BibitemOpen
  \bibfield  {author} {\bibinfo {author} {\bibnamefont {{The Extreme Light
  Infrastructure (ELI)}}},\ }\href@noop {} {}\bibinfo {howpublished}
  {\url{http://www.eli-laser.eu/}}\BibitemShut {NoStop}%
\bibitem [{\citenamefont {{Exawatt Center for Extreme Light Stidies
  (XCELS)}}()}]{XCELS}%
  \BibitemOpen
  \bibfield  {author} {\bibinfo {author} {\bibnamefont {{Exawatt Center for
  Extreme Light Stidies (XCELS)}}},\ }\href@noop {} {}\bibinfo {howpublished}
  {\url{http://www.xcels.iapras.ru/}}\BibitemShut {NoStop}%
\bibitem [{\citenamefont {Marklund}\ and\ \citenamefont
  {Shukla}(2006)}]{Marklund_2006}%
  \BibitemOpen
  \bibfield  {author} {\bibinfo {author} {\bibfnamefont {M.}~\bibnamefont
  {Marklund}}\ and\ \bibinfo {author} {\bibfnamefont {P.~K.}\ \bibnamefont
  {Shukla}},\ }\href@noop {} {\bibfield  {journal} {\bibinfo  {journal} {Rev.
  Mod. Phys.}\ }\textbf {\bibinfo {volume} {78}},\ \bibinfo {pages} {591}
  (\bibinfo {year} {2006})}\BibitemShut {NoStop}%
\bibitem [{\citenamefont {Mourou}\ \emph {et~al.}(2006)\citenamefont {Mourou},
  \citenamefont {Tajima},\ and\ \citenamefont {Bulanov}}]{Mourou_2006}%
  \BibitemOpen
  \bibfield  {author} {\bibinfo {author} {\bibfnamefont {G.~A.}\ \bibnamefont
  {Mourou}}, \bibinfo {author} {\bibfnamefont {T.}~\bibnamefont {Tajima}}, \
  and\ \bibinfo {author} {\bibfnamefont {S.~V.}\ \bibnamefont {Bulanov}},\
  }\href@noop {} {\bibfield  {journal} {\bibinfo  {journal} {Rev. Mod. Phys.}\
  }\textbf {\bibinfo {volume} {78}},\ \bibinfo {pages} {309} (\bibinfo {year}
  {2006})}\BibitemShut {NoStop}%
\bibitem [{\citenamefont {Esarey}\ \emph {et~al.}(2009)\citenamefont {Esarey},
  \citenamefont {Schroeder},\ and\ \citenamefont {Leemans}}]{Esarey_2009}%
  \BibitemOpen
  \bibfield  {author} {\bibinfo {author} {\bibfnamefont {E.}~\bibnamefont
  {Esarey}}, \bibinfo {author} {\bibfnamefont {C.~B.}\ \bibnamefont
  {Schroeder}}, \ and\ \bibinfo {author} {\bibfnamefont {W.~P.}\ \bibnamefont
  {Leemans}},\ }\href@noop {} {\bibfield  {journal} {\bibinfo  {journal} {Rev.
  Mod. Phys.}\ }\textbf {\bibinfo {volume} {81}},\ \bibinfo {pages} {1229}
  (\bibinfo {year} {2009})}\BibitemShut {NoStop}%
\bibitem [{\citenamefont {{Di Piazza}}\ \emph {et~al.}(2012)\citenamefont {{Di
  Piazza}}, \citenamefont {M\"uller}, \citenamefont {Hatsagortsyan},\ and\
  \citenamefont {Keitel}}]{RMP_2012}%
  \BibitemOpen
  \bibfield  {author} {\bibinfo {author} {\bibfnamefont {A.}~\bibnamefont {{Di
  Piazza}}}, \bibinfo {author} {\bibfnamefont {C.}~\bibnamefont {M\"uller}},
  \bibinfo {author} {\bibfnamefont {K.~Z.}\ \bibnamefont {Hatsagortsyan}}, \
  and\ \bibinfo {author} {\bibfnamefont {C.~H.}\ \bibnamefont {Keitel}},\
  }\href@noop {} {\bibfield  {journal} {\bibinfo  {journal} {Rev. Mod. Phys.}\
  }\textbf {\bibinfo {volume} {84}},\ \bibinfo {pages} {1177} (\bibinfo {year}
  {2012})}\BibitemShut {NoStop}%
\bibitem [{\citenamefont {Mourou}\ and\ \citenamefont
  {Tajima}(2014)}]{Mourou_2014}%
  \BibitemOpen
  \bibfield  {author} {\bibinfo {author} {\bibfnamefont {G.}~\bibnamefont
  {Mourou}}\ and\ \bibinfo {author} {\bibfnamefont {T.}~\bibnamefont
  {Tajima}},\ }\href@noop {} {\bibfield  {journal} {\bibinfo  {journal} {Eur.
  Phys. J. Spec. Topics}\ }\textbf {\bibinfo {volume} {223}},\ \bibinfo {pages}
  {979} (\bibinfo {year} {2014})}\BibitemShut {NoStop}%
\bibitem [{\citenamefont {Meuren}\ and\ \citenamefont
  {Di~Piazza}(2011)}]{Meuren_2011}%
  \BibitemOpen
  \bibfield  {author} {\bibinfo {author} {\bibfnamefont {S.}~\bibnamefont
  {Meuren}}\ and\ \bibinfo {author} {\bibfnamefont {A.}~\bibnamefont
  {Di~Piazza}},\ }\href@noop {} {\bibfield  {journal} {\bibinfo  {journal}
  {Phys. Rev. Lett.}\ }\textbf {\bibinfo {volume} {107}},\ \bibinfo {pages}
  {260401} (\bibinfo {year} {2011})}\BibitemShut {NoStop}%
\bibitem [{\citenamefont {Boca}\ and\ \citenamefont
  {Florescu}(2009)}]{Boca_2009}%
  \BibitemOpen
  \bibfield  {author} {\bibinfo {author} {\bibfnamefont {M.}~\bibnamefont
  {Boca}}\ and\ \bibinfo {author} {\bibfnamefont {V.}~\bibnamefont
  {Florescu}},\ }\href@noop {} {\bibfield  {journal} {\bibinfo  {journal}
  {Phys. Rev. A}\ }\textbf {\bibinfo {volume} {80}},\ \bibinfo {pages} {053403}
  (\bibinfo {year} {2009})}\BibitemShut {NoStop}%
\bibitem [{\citenamefont {Mackenroth}\ \emph {et~al.}(2010)\citenamefont
  {Mackenroth}, \citenamefont {Di~Piazza},\ and\ \citenamefont
  {Keitel}}]{Mackenroth_2010}%
  \BibitemOpen
  \bibfield  {author} {\bibinfo {author} {\bibfnamefont {F.}~\bibnamefont
  {Mackenroth}}, \bibinfo {author} {\bibfnamefont {A.}~\bibnamefont
  {Di~Piazza}}, \ and\ \bibinfo {author} {\bibfnamefont {C.~H.}\ \bibnamefont
  {Keitel}},\ }\href {\doibase 10.1103/PhysRevLett.105.063903} {\bibfield
  {journal} {\bibinfo  {journal} {Phys. Rev. Lett.}\ }\textbf {\bibinfo
  {volume} {105}},\ \bibinfo {pages} {063903} (\bibinfo {year}
  {2010})}\BibitemShut {NoStop}%
\bibitem [{\citenamefont {Seipt}\ and\ \citenamefont
  {K\"ampfer}(2013)}]{Seipt_2013}%
  \BibitemOpen
  \bibfield  {author} {\bibinfo {author} {\bibfnamefont {D.}~\bibnamefont
  {Seipt}}\ and\ \bibinfo {author} {\bibfnamefont {B.}~\bibnamefont
  {K\"ampfer}},\ }\href@noop {} {\bibfield  {journal} {\bibinfo  {journal}
  {Phys. Rev. A}\ }\textbf {\bibinfo {volume} {88}},\ \bibinfo {pages} {012127}
  (\bibinfo {year} {2013})}\BibitemShut {NoStop}%
\bibitem [{\citenamefont {Hebenstreit}\ \emph {et~al.}(2009)\citenamefont
  {Hebenstreit}, \citenamefont {Alkofer}, \citenamefont {Dunne},\ and\
  \citenamefont {Gies}}]{Hebenstreit_2009}%
  \BibitemOpen
  \bibfield  {author} {\bibinfo {author} {\bibfnamefont {F.}~\bibnamefont
  {Hebenstreit}}, \bibinfo {author} {\bibfnamefont {R.}~\bibnamefont
  {Alkofer}}, \bibinfo {author} {\bibfnamefont {G.~V.}\ \bibnamefont {Dunne}},
  \ and\ \bibinfo {author} {\bibfnamefont {H.}~\bibnamefont {Gies}},\
  }\href@noop {} {\bibfield  {journal} {\bibinfo  {journal} {Phys. Rev. Lett.}\
  }\textbf {\bibinfo {volume} {102}},\ \bibinfo {pages} {150404} (\bibinfo
  {year} {2009})}\BibitemShut {NoStop}%
\bibitem [{\citenamefont {Krajewska}\ and\ \citenamefont
  {Kami\ifmmode~\acute{n}\else \'{n}\fi{}ski}(2012)}]{Krajewska_2012}%
  \BibitemOpen
  \bibfield  {author} {\bibinfo {author} {\bibfnamefont {K.}~\bibnamefont
  {Krajewska}}\ and\ \bibinfo {author} {\bibfnamefont {J.~Z.}\ \bibnamefont
  {Kami\ifmmode~\acute{n}\else \'{n}\fi{}ski}},\ }\href@noop {} {\bibfield
  {journal} {\bibinfo  {journal} {Phys. Rev. A}\ }\textbf {\bibinfo {volume}
  {86}},\ \bibinfo {pages} {052104} (\bibinfo {year} {2012})}\BibitemShut
  {NoStop}%
\bibitem [{\citenamefont {Abdukerim}\ \emph {et~al.}(2013)\citenamefont
  {Abdukerim}, \citenamefont {Li},\ and\ \citenamefont {Xie}}]{Nuriman_2013}%
  \BibitemOpen
  \bibfield  {author} {\bibinfo {author} {\bibfnamefont {N.}~\bibnamefont
  {Abdukerim}}, \bibinfo {author} {\bibfnamefont {Z.-L.}\ \bibnamefont {Li}}, \
  and\ \bibinfo {author} {\bibfnamefont {B.-S.}\ \bibnamefont {Xie}},\
  }\href@noop {} {\bibfield  {journal} {\bibinfo  {journal} {Phys. Lett. B}\
  }\textbf {\bibinfo {volume} {726}},\ \bibinfo {pages} {820} (\bibinfo {year}
  {2013})}\BibitemShut {NoStop}%
\bibitem [{\citenamefont {Titov}\ \emph {et~al.}(2016)\citenamefont {Titov},
  \citenamefont {K\"ampfer}, \citenamefont {Hosaka}, \citenamefont {Nousch},\
  and\ \citenamefont {Seipt}}]{Titov_2016}%
  \BibitemOpen
  \bibfield  {author} {\bibinfo {author} {\bibfnamefont {A.~I.}\ \bibnamefont
  {Titov}}, \bibinfo {author} {\bibfnamefont {B.}~\bibnamefont {K\"ampfer}},
  \bibinfo {author} {\bibfnamefont {A.}~\bibnamefont {Hosaka}}, \bibinfo
  {author} {\bibfnamefont {T.}~\bibnamefont {Nousch}}, \ and\ \bibinfo {author}
  {\bibfnamefont {D.}~\bibnamefont {Seipt}},\ }\href@noop {} {\bibfield
  {journal} {\bibinfo  {journal} {Phys. Rev. D}\ }\textbf {\bibinfo {volume}
  {93}},\ \bibinfo {pages} {045010} (\bibinfo {year} {2016})}\BibitemShut
  {NoStop}%
\bibitem [{\citenamefont {Jansen}\ and\ \citenamefont
  {M\"uller}(2016)}]{Jansen_2016}%
  \BibitemOpen
  \bibfield  {author} {\bibinfo {author} {\bibfnamefont {M.~J.~A.}\
  \bibnamefont {Jansen}}\ and\ \bibinfo {author} {\bibfnamefont
  {C.}~\bibnamefont {M\"uller}},\ }\href@noop {} {\bibfield  {journal}
  {\bibinfo  {journal} {Phys. Rev. D}\ }\textbf {\bibinfo {volume} {93}},\
  \bibinfo {pages} {053011} (\bibinfo {year} {2016})}\BibitemShut {NoStop}%
\bibitem [{\citenamefont {Meuren}\ \emph {et~al.}(2016)\citenamefont {Meuren},
  \citenamefont {Keitel},\ and\ \citenamefont {Di~Piazza}}]{Meuren_2016}%
  \BibitemOpen
  \bibfield  {author} {\bibinfo {author} {\bibfnamefont {S.}~\bibnamefont
  {Meuren}}, \bibinfo {author} {\bibfnamefont {C.~H.}\ \bibnamefont {Keitel}},
  \ and\ \bibinfo {author} {\bibfnamefont {A.}~\bibnamefont {Di~Piazza}},\
  }\href@noop {} {\bibfield  {journal} {\bibinfo  {journal} {Phys. Rev. D}\
  }\textbf {\bibinfo {volume} {93}},\ \bibinfo {pages} {085028} (\bibinfo
  {year} {2016})}\BibitemShut {NoStop}%
\bibitem [{\citenamefont {Paulus}\ \emph {et~al.}(2001)\citenamefont {Paulus},
  \citenamefont {Grasbon}, \citenamefont {Walther}, \citenamefont {Villoresi},
  \citenamefont {Nisoli}, \citenamefont {Stagira}, \citenamefont {Priori},\
  and\ \citenamefont {{De Silvestri}}}]{paulus_2001}%
  \BibitemOpen
  \bibfield  {author} {\bibinfo {author} {\bibfnamefont {G.~G.}\ \bibnamefont
  {Paulus}}, \bibinfo {author} {\bibfnamefont {F.}~\bibnamefont {Grasbon}},
  \bibinfo {author} {\bibfnamefont {H.}~\bibnamefont {Walther}}, \bibinfo
  {author} {\bibfnamefont {P.}~\bibnamefont {Villoresi}}, \bibinfo {author}
  {\bibfnamefont {M.}~\bibnamefont {Nisoli}}, \bibinfo {author} {\bibfnamefont
  {S.}~\bibnamefont {Stagira}}, \bibinfo {author} {\bibfnamefont
  {E.}~\bibnamefont {Priori}}, \ and\ \bibinfo {author} {\bibfnamefont
  {S.}~\bibnamefont {{De Silvestri}}},\ }\href@noop {} {\bibfield  {journal}
  {\bibinfo  {journal} {Nature}\ }\textbf {\bibinfo {volume} {414}},\ \bibinfo
  {pages} {182} (\bibinfo {year} {2001})}\BibitemShut {NoStop}%
\bibitem [{\citenamefont {Paulus}\ \emph {et~al.}(2003)\citenamefont {Paulus},
  \citenamefont {Lindner}, \citenamefont {Walther}, \citenamefont
  {Baltu\ifmmode~\check{s}\else \v{s}\fi{}ka}, \citenamefont {Goulielmakis},
  \citenamefont {Lezius},\ and\ \citenamefont {Krausz}}]{Paulus_2003}%
  \BibitemOpen
  \bibfield  {author} {\bibinfo {author} {\bibfnamefont {G.~G.}\ \bibnamefont
  {Paulus}}, \bibinfo {author} {\bibfnamefont {F.}~\bibnamefont {Lindner}},
  \bibinfo {author} {\bibfnamefont {H.}~\bibnamefont {Walther}}, \bibinfo
  {author} {\bibfnamefont {A.}~\bibnamefont {Baltu\ifmmode~\check{s}\else
  \v{s}\fi{}ka}}, \bibinfo {author} {\bibfnamefont {E.}~\bibnamefont
  {Goulielmakis}}, \bibinfo {author} {\bibfnamefont {M.}~\bibnamefont
  {Lezius}}, \ and\ \bibinfo {author} {\bibfnamefont {F.}~\bibnamefont
  {Krausz}},\ }\href@noop {} {\bibfield  {journal} {\bibinfo  {journal} {Phys.
  Rev. Lett.}\ }\textbf {\bibinfo {volume} {91}},\ \bibinfo {pages} {253004}
  (\bibinfo {year} {2003})}\BibitemShut {NoStop}%
\bibitem [{\citenamefont {Wittmann}\ \emph {et~al.}(2009)\citenamefont
  {Wittmann}, \citenamefont {Horvath}, \citenamefont {Helml}, \citenamefont
  {Schatzel}, \citenamefont {Gu}, \citenamefont {Cavalieri}, \citenamefont
  {Paulus},\ and\ \citenamefont {Kienberger}}]{Wittmann_2009}%
  \BibitemOpen
  \bibfield  {author} {\bibinfo {author} {\bibfnamefont {T.}~\bibnamefont
  {Wittmann}}, \bibinfo {author} {\bibfnamefont {B.}~\bibnamefont {Horvath}},
  \bibinfo {author} {\bibfnamefont {W.}~\bibnamefont {Helml}}, \bibinfo
  {author} {\bibfnamefont {M.~G.}\ \bibnamefont {Schatzel}}, \bibinfo {author}
  {\bibfnamefont {X.}~\bibnamefont {Gu}}, \bibinfo {author} {\bibfnamefont
  {A.~L.}\ \bibnamefont {Cavalieri}}, \bibinfo {author} {\bibfnamefont {G.~G.}\
  \bibnamefont {Paulus}}, \ and\ \bibinfo {author} {\bibfnamefont
  {R.}~\bibnamefont {Kienberger}},\ }\href@noop {} {\bibfield  {journal}
  {\bibinfo  {journal} {Nat. Phys.}\ }\textbf {\bibinfo {volume} {5}},\
  \bibinfo {pages} {357} (\bibinfo {year} {2009})}\BibitemShut {NoStop}%
\bibitem [{\citenamefont {Goulielmakis}\ \emph {et~al.}(2004)\citenamefont
  {Goulielmakis}, \citenamefont {Uiberacker}, \citenamefont {Kienberger},
  \citenamefont {Baltuska}, \citenamefont {Yakovlev}, \citenamefont {Scrinzi},
  \citenamefont {Westerwalbesloh}, \citenamefont {Kleineberg}, \citenamefont
  {Heinzmann}, \citenamefont {Drescher},\ and\ \citenamefont
  {Krausz}}]{Goulielmakis_2004}%
  \BibitemOpen
  \bibfield  {author} {\bibinfo {author} {\bibfnamefont {E.}~\bibnamefont
  {Goulielmakis}}, \bibinfo {author} {\bibfnamefont {M.}~\bibnamefont
  {Uiberacker}}, \bibinfo {author} {\bibfnamefont {R.}~\bibnamefont
  {Kienberger}}, \bibinfo {author} {\bibfnamefont {A.}~\bibnamefont
  {Baltuska}}, \bibinfo {author} {\bibfnamefont {V.}~\bibnamefont {Yakovlev}},
  \bibinfo {author} {\bibfnamefont {A.}~\bibnamefont {Scrinzi}}, \bibinfo
  {author} {\bibfnamefont {T.}~\bibnamefont {Westerwalbesloh}}, \bibinfo
  {author} {\bibfnamefont {U.}~\bibnamefont {Kleineberg}}, \bibinfo {author}
  {\bibfnamefont {U.}~\bibnamefont {Heinzmann}}, \bibinfo {author}
  {\bibfnamefont {M.}~\bibnamefont {Drescher}}, \ and\ \bibinfo {author}
  {\bibfnamefont {F.}~\bibnamefont {Krausz}},\ }\href@noop {} {\bibfield
  {journal} {\bibinfo  {journal} {Science}\ }\textbf {\bibinfo {volume}
  {305}},\ \bibinfo {pages} {1267} (\bibinfo {year} {2004})}\BibitemShut
  {NoStop}%
\bibitem [{\citenamefont {Kre\ss}\ \emph {et~al.}(2006)\citenamefont {Kre\ss},
  \citenamefont {L\"offler}, \citenamefont {Thomson}, \citenamefont {D\"orner},
  \citenamefont {Gimpel}, \citenamefont {Zrost}, \citenamefont {Ergler},
  \citenamefont {Moshammer}, \citenamefont {Morgner}, \citenamefont {Ullrich},\
  and\ \citenamefont {Roskos}}]{Kress_2006}%
  \BibitemOpen
  \bibfield  {author} {\bibinfo {author} {\bibfnamefont {M.}~\bibnamefont
  {Kre\ss}}, \bibinfo {author} {\bibfnamefont {T.}~\bibnamefont {L\"offler}},
  \bibinfo {author} {\bibfnamefont {M.~D.}\ \bibnamefont {Thomson}}, \bibinfo
  {author} {\bibfnamefont {R.}~\bibnamefont {D\"orner}}, \bibinfo {author}
  {\bibfnamefont {H.}~\bibnamefont {Gimpel}}, \bibinfo {author} {\bibfnamefont
  {K.}~\bibnamefont {Zrost}}, \bibinfo {author} {\bibfnamefont
  {T.}~\bibnamefont {Ergler}}, \bibinfo {author} {\bibfnamefont
  {R.}~\bibnamefont {Moshammer}}, \bibinfo {author} {\bibfnamefont
  {U.}~\bibnamefont {Morgner}}, \bibinfo {author} {\bibfnamefont
  {J.}~\bibnamefont {Ullrich}}, \ and\ \bibinfo {author} {\bibfnamefont
  {H.~G.}\ \bibnamefont {Roskos}},\ }\href@noop {} {\bibfield  {journal}
  {\bibinfo  {journal} {Nat. Phys.}\ }\textbf {\bibinfo {volume} {2}},\
  \bibinfo {pages} {327} (\bibinfo {year} {2006})}\BibitemShut {NoStop}%
\bibitem [{\citenamefont {{Di~Piazza}}\ \emph {et~al.}(2009)\citenamefont
  {{Di~Piazza}}, \citenamefont {Hatsagortsyan},\ and\ \citenamefont
  {Keitel}}]{DiPiazza_2009}%
  \BibitemOpen
  \bibfield  {author} {\bibinfo {author} {\bibfnamefont {A.}~\bibnamefont
  {{Di~Piazza}}}, \bibinfo {author} {\bibfnamefont {K.~Z.}\ \bibnamefont
  {Hatsagortsyan}}, \ and\ \bibinfo {author} {\bibfnamefont {C.~H.}\
  \bibnamefont {Keitel}},\ }\href@noop {} {\bibfield  {journal} {\bibinfo
  {journal} {Phys. Rev. Lett.}\ }\textbf {\bibinfo {volume} {102}},\ \bibinfo
  {pages} {254802} (\bibinfo {year} {2009})}\BibitemShut {NoStop}%
\bibitem [{\citenamefont {Li}\ \emph {et~al.}(2017)\citenamefont {Li},
  \citenamefont {Chen}, \citenamefont {Hatsagortsyan},\ and\ \citenamefont
  {Keitel}}]{Jianxing_2017}%
  \BibitemOpen
  \bibfield  {author} {\bibinfo {author} {\bibfnamefont {J.-X.}\ \bibnamefont
  {Li}}, \bibinfo {author} {\bibfnamefont {Y.-Y.}\ \bibnamefont {Chen}},
  \bibinfo {author} {\bibfnamefont {K.~Z.}\ \bibnamefont {Hatsagortsyan}}, \
  and\ \bibinfo {author} {\bibfnamefont {C.~H.}\ \bibnamefont {Keitel}},\
  }\href@noop {} {\bibfield  {journal} {\bibinfo  {journal} {arXiv:1612.06796}\
  } (\bibinfo {year} {2017})}\BibitemShut {NoStop}%
\bibitem [{\citenamefont {Shen}\ and\ \citenamefont {White}(1972)}]{Shen_1972}%
  \BibitemOpen
  \bibfield  {author} {\bibinfo {author} {\bibfnamefont {C.~S.}\ \bibnamefont
  {Shen}}\ and\ \bibinfo {author} {\bibfnamefont {D.}~\bibnamefont {White}},\
  }\href@noop {} {\bibfield  {journal} {\bibinfo  {journal} {Phys. Rev. Lett.}\
  }\textbf {\bibinfo {volume} {28}},\ \bibinfo {pages} {455} (\bibinfo {year}
  {1972})}\BibitemShut {NoStop}%
\bibitem [{\citenamefont {Duclous}\ \emph {et~al.}(2011)\citenamefont
  {Duclous}, \citenamefont {Kirk},\ and\ \citenamefont {Bell}}]{Duclous_2011}%
  \BibitemOpen
  \bibfield  {author} {\bibinfo {author} {\bibfnamefont {R.}~\bibnamefont
  {Duclous}}, \bibinfo {author} {\bibfnamefont {J.~G.}\ \bibnamefont {Kirk}}, \
  and\ \bibinfo {author} {\bibfnamefont {A.~R.}\ \bibnamefont {Bell}},\
  }\href@noop {} {\bibfield  {journal} {\bibinfo  {journal} {Plasma Phys.
  Contr. F.}\ }\textbf {\bibinfo {volume} {53}},\ \bibinfo {pages} {015009}
  (\bibinfo {year} {2011})}\BibitemShut {NoStop}%
\bibitem [{\citenamefont {Neitz}\ and\ \citenamefont
  {Di~Piazza}(2013)}]{Neitz_2013}%
  \BibitemOpen
  \bibfield  {author} {\bibinfo {author} {\bibfnamefont {N.}~\bibnamefont
  {Neitz}}\ and\ \bibinfo {author} {\bibfnamefont {A.}~\bibnamefont
  {Di~Piazza}},\ }\href@noop {} {\bibfield  {journal} {\bibinfo  {journal}
  {Phys. Rev. Lett.}\ }\textbf {\bibinfo {volume} {111}},\ \bibinfo {pages}
  {054802} (\bibinfo {year} {2013})}\BibitemShut {NoStop}%
\bibitem [{\citenamefont {Reiss}(1962)}]{Reiss_1962}%
  \BibitemOpen
  \bibfield  {author} {\bibinfo {author} {\bibfnamefont {H.~R.}\ \bibnamefont
  {Reiss}},\ }\href@noop {} {\bibfield  {journal} {\bibinfo  {journal} {J.
  Math. Phys. (N.Y.)}\ }\textbf {\bibinfo {volume} {3}},\ \bibinfo {pages} {59}
  (\bibinfo {year} {1962})}\BibitemShut {NoStop}%
\bibitem [{\citenamefont {Ritus}(1985)}]{Ritus_1985}%
  \BibitemOpen
  \bibfield  {author} {\bibinfo {author} {\bibfnamefont {V.~I.}\ \bibnamefont
  {Ritus}},\ }\href@noop {} {\bibfield  {journal} {\bibinfo  {journal} {J. Sov.
  Laser Res.}\ }\textbf {\bibinfo {volume} {6}},\ \bibinfo {pages} {497}
  (\bibinfo {year} {1985})}\BibitemShut {NoStop}%
\bibitem [{\citenamefont {Bula}\ \emph {et~al.}(1996)\citenamefont {Bula},
  \citenamefont {McDonald}, \citenamefont {Prebys}, \citenamefont {Bamber},
  \citenamefont {Boege}, \citenamefont {Kotseroglou}, \citenamefont
  {Melissinos}, \citenamefont {Meyerhofer}, \citenamefont {Ragg}, \citenamefont
  {Burke}, \citenamefont {Field}, \citenamefont {Horton-Smith}, \citenamefont
  {Odian}, \citenamefont {Spencer}, \citenamefont {Walz}, \citenamefont
  {Berridge}, \citenamefont {Bugg}, \citenamefont {Shmakov},\ and\
  \citenamefont {Weidemann}}]{Bula_1996}%
  \BibitemOpen
  \bibfield  {author} {\bibinfo {author} {\bibfnamefont {C.}~\bibnamefont
  {Bula}}, \bibinfo {author} {\bibfnamefont {K.~T.}\ \bibnamefont {McDonald}},
  \bibinfo {author} {\bibfnamefont {E.~J.}\ \bibnamefont {Prebys}}, \bibinfo
  {author} {\bibfnamefont {C.}~\bibnamefont {Bamber}}, \bibinfo {author}
  {\bibfnamefont {S.}~\bibnamefont {Boege}}, \bibinfo {author} {\bibfnamefont
  {T.}~\bibnamefont {Kotseroglou}}, \bibinfo {author} {\bibfnamefont {A.~C.}\
  \bibnamefont {Melissinos}}, \bibinfo {author} {\bibfnamefont {D.~D.}\
  \bibnamefont {Meyerhofer}}, \bibinfo {author} {\bibfnamefont
  {W.}~\bibnamefont {Ragg}}, \bibinfo {author} {\bibfnamefont {D.~L.}\
  \bibnamefont {Burke}}, \bibinfo {author} {\bibfnamefont {R.~C.}\ \bibnamefont
  {Field}}, \bibinfo {author} {\bibfnamefont {G.}~\bibnamefont {Horton-Smith}},
  \bibinfo {author} {\bibfnamefont {A.~C.}\ \bibnamefont {Odian}}, \bibinfo
  {author} {\bibfnamefont {J.~E.}\ \bibnamefont {Spencer}}, \bibinfo {author}
  {\bibfnamefont {D.}~\bibnamefont {Walz}}, \bibinfo {author} {\bibfnamefont
  {S.~C.}\ \bibnamefont {Berridge}}, \bibinfo {author} {\bibfnamefont {W.~M.}\
  \bibnamefont {Bugg}}, \bibinfo {author} {\bibfnamefont {K.}~\bibnamefont
  {Shmakov}}, \ and\ \bibinfo {author} {\bibfnamefont {A.~W.}\ \bibnamefont
  {Weidemann}},\ }\href@noop {} {\bibfield  {journal} {\bibinfo  {journal}
  {Phys. Rev. Lett.}\ }\textbf {\bibinfo {volume} {76}},\ \bibinfo {pages}
  {3116} (\bibinfo {year} {1996})}\BibitemShut {NoStop}%
\bibitem [{\citenamefont {Panek}\ \emph {et~al.}(2002)\citenamefont {Panek},
  \citenamefont {Kami\ifmmode~\acute{n}\else \'{n}\fi{}ski},\ and\
  \citenamefont {Ehlotzky}}]{Panek_2002}%
  \BibitemOpen
  \bibfield  {author} {\bibinfo {author} {\bibfnamefont {P.}~\bibnamefont
  {Panek}}, \bibinfo {author} {\bibfnamefont {J.~Z.}\ \bibnamefont
  {Kami\ifmmode~\acute{n}\else \'{n}\fi{}ski}}, \ and\ \bibinfo {author}
  {\bibfnamefont {F.}~\bibnamefont {Ehlotzky}},\ }\href@noop {} {\bibfield
  {journal} {\bibinfo  {journal} {Phys. Rev. A}\ }\textbf {\bibinfo {volume}
  {65}},\ \bibinfo {pages} {033408} (\bibinfo {year} {2002})}\BibitemShut
  {NoStop}%
\bibitem [{\citenamefont {Seipt}\ and\ \citenamefont
  {K\"ampfer}(2011)}]{Seipt_2011}%
  \BibitemOpen
  \bibfield  {author} {\bibinfo {author} {\bibfnamefont {D.}~\bibnamefont
  {Seipt}}\ and\ \bibinfo {author} {\bibfnamefont {B.}~\bibnamefont
  {K\"ampfer}},\ }\href {\doibase 10.1103/PhysRevA.83.022101} {\bibfield
  {journal} {\bibinfo  {journal} {Phys. Rev. A}\ }\textbf {\bibinfo {volume}
  {83}},\ \bibinfo {pages} {022101} (\bibinfo {year} {2011})}\BibitemShut
  {NoStop}%
\bibitem [{\citenamefont {Vranic}\ \emph {et~al.}(2016)\citenamefont {Vranic},
  \citenamefont {Grismayer}, \citenamefont {Fonseca},\ and\ \citenamefont
  {Silva}}]{Vranic_2016}%
  \BibitemOpen
  \bibfield  {author} {\bibinfo {author} {\bibfnamefont {M.}~\bibnamefont
  {Vranic}}, \bibinfo {author} {\bibfnamefont {T.}~\bibnamefont {Grismayer}},
  \bibinfo {author} {\bibfnamefont {R.~A.}\ \bibnamefont {Fonseca}}, \ and\
  \bibinfo {author} {\bibfnamefont {L.~O.}\ \bibnamefont {Silva}},\ }\href@noop
  {} {\bibfield  {journal} {\bibinfo  {journal} {New J.Phys.}\ }\textbf
  {\bibinfo {volume} {18}},\ \bibinfo {pages} {073035} (\bibinfo {year}
  {2016})}\BibitemShut {NoStop}%
\bibitem [{\citenamefont {Li}\ \emph {et~al.}(2015)\citenamefont {Li},
  \citenamefont {Hatsagortsyan}, \citenamefont {Galow},\ and\ \citenamefont
  {Keitel}}]{Li_2015}%
  \BibitemOpen
  \bibfield  {author} {\bibinfo {author} {\bibfnamefont {J.-X.}\ \bibnamefont
  {Li}}, \bibinfo {author} {\bibfnamefont {K.~Z.}\ \bibnamefont
  {Hatsagortsyan}}, \bibinfo {author} {\bibfnamefont {B.~J.}\ \bibnamefont
  {Galow}}, \ and\ \bibinfo {author} {\bibfnamefont {C.~H.}\ \bibnamefont
  {Keitel}},\ }\href@noop {} {\bibfield  {journal} {\bibinfo  {journal} {Phys.
  Rev. Lett.}\ }\textbf {\bibinfo {volume} {115}},\ \bibinfo {pages} {204801}
  (\bibinfo {year} {2015})}\BibitemShut {NoStop}%
\bibitem [{\citenamefont {Elkina}\ \emph {et~al.}(2011)\citenamefont {Elkina},
  \citenamefont {Fedotov}, \citenamefont {Kostyukov}, \citenamefont {Legkov},
  \citenamefont {Narozhny}, \citenamefont {Nerush},\ and\ \citenamefont
  {Ruhl}}]{Elkina_2011}%
  \BibitemOpen
  \bibfield  {author} {\bibinfo {author} {\bibfnamefont {N.~V.}\ \bibnamefont
  {Elkina}}, \bibinfo {author} {\bibfnamefont {A.~M.}\ \bibnamefont {Fedotov}},
  \bibinfo {author} {\bibfnamefont {I.~Y.}\ \bibnamefont {Kostyukov}}, \bibinfo
  {author} {\bibfnamefont {M.~V.}\ \bibnamefont {Legkov}}, \bibinfo {author}
  {\bibfnamefont {N.~B.}\ \bibnamefont {Narozhny}}, \bibinfo {author}
  {\bibfnamefont {E.~N.}\ \bibnamefont {Nerush}}, \ and\ \bibinfo {author}
  {\bibfnamefont {H.}~\bibnamefont {Ruhl}},\ }\href@noop {} {\bibfield
  {journal} {\bibinfo  {journal} {Phys. Rev. ST Accel. Beams}\ }\textbf
  {\bibinfo {volume} {14}},\ \bibinfo {pages} {054401} (\bibinfo {year}
  {2011})}\BibitemShut {NoStop}%
\bibitem [{\citenamefont {Ridgers}\ \emph {et~al.}(2014)\citenamefont
  {Ridgers}, \citenamefont {Kirk}, \citenamefont {Duclous}, \citenamefont
  {Blackburn}, \citenamefont {Brady}, \citenamefont {Bennett}, \citenamefont
  {Arber},\ and\ \citenamefont {Bell}}]{Ridgers_2014}%
  \BibitemOpen
  \bibfield  {author} {\bibinfo {author} {\bibfnamefont {C.~P.}\ \bibnamefont
  {Ridgers}}, \bibinfo {author} {\bibfnamefont {J.~G.}\ \bibnamefont {Kirk}},
  \bibinfo {author} {\bibfnamefont {R.}~\bibnamefont {Duclous}}, \bibinfo
  {author} {\bibfnamefont {T.~G.}\ \bibnamefont {Blackburn}}, \bibinfo {author}
  {\bibfnamefont {C.~S.}\ \bibnamefont {Brady}}, \bibinfo {author}
  {\bibfnamefont {K.}~\bibnamefont {Bennett}}, \bibinfo {author} {\bibfnamefont
  {T.~D.}\ \bibnamefont {Arber}}, \ and\ \bibinfo {author} {\bibfnamefont
  {A.~R.}\ \bibnamefont {Bell}},\ }\href@noop {} {\bibfield  {journal}
  {\bibinfo  {journal} {J. Compt. Phys.}\ }\textbf {\bibinfo {volume} {260}},\
  \bibinfo {pages} {273} (\bibinfo {year} {2014})}\BibitemShut {NoStop}%
\bibitem [{\citenamefont {Green}\ and\ \citenamefont
  {Harvey}(2015)}]{Green_2015}%
  \BibitemOpen
  \bibfield  {author} {\bibinfo {author} {\bibfnamefont {D.~G.}\ \bibnamefont
  {Green}}\ and\ \bibinfo {author} {\bibfnamefont {C.~N.}\ \bibnamefont
  {Harvey}},\ }\href@noop {} {\bibfield  {journal} {\bibinfo  {journal} {Comp.
  Phys. Commun.}\ }\textbf {\bibinfo {volume} {192}},\ \bibinfo {pages} {313}
  (\bibinfo {year} {2015})}\BibitemShut {NoStop}%
\bibitem [{Sup()}]{Suppl_material}%
  \BibitemOpen
  \href@noop {} {}\bibinfo {howpublished} {See the Supplemental Materials for
  the details, which includes the employed electromegnetic fields of the laser
  pulse, and the description of the applied methods for the calculation of the
  electron radiation.}\BibitemShut {Stop}%
\bibitem [{\citenamefont {Baier}\ \emph {et~al.}(1994)\citenamefont {Baier},
  \citenamefont {Katkov},\ and\ \citenamefont {Strakhovenko}}]{Baier_b_1994}%
  \BibitemOpen
  \bibfield  {author} {\bibinfo {author} {\bibfnamefont {V.~N.}\ \bibnamefont
  {Baier}}, \bibinfo {author} {\bibfnamefont {V.~M.}\ \bibnamefont {Katkov}}, \
  and\ \bibinfo {author} {\bibfnamefont {V.~M.}\ \bibnamefont {Strakhovenko}},\
  }\href@noop {} {\emph {\bibinfo {title} {Electromagnetic Processes at High
  Energies in Oriented Single Crystals}}}\ (\bibinfo  {publisher} {World
  Scientific, Singapore},\ \bibinfo {year} {1994})\BibitemShut {NoStop}%
\bibitem [{\citenamefont {Salamin}\ and\ \citenamefont
  {Keitel}(2002)}]{Salamin_2002}%
  \BibitemOpen
  \bibfield  {author} {\bibinfo {author} {\bibfnamefont {Y.~I.}\ \bibnamefont
  {Salamin}}\ and\ \bibinfo {author} {\bibfnamefont {C.~H.}\ \bibnamefont
  {Keitel}},\ }\href@noop {} {\bibfield  {journal} {\bibinfo  {journal} {Phys.
  Rev. Lett.}\ }\textbf {\bibinfo {volume} {88}},\ \bibinfo {pages} {095005}
  (\bibinfo {year} {2002})}\BibitemShut {NoStop}%
\bibitem [{\citenamefont {Rousse}\ \emph {et~al.}(2004)\citenamefont {Rousse},
  \citenamefont {Phuoc}, \citenamefont {Shah}, \citenamefont {Pukhov},
  \citenamefont {Lefebvre}, \citenamefont {Malka}, \citenamefont {Kiselev},
  \citenamefont {Burgy}, \citenamefont {Rousseau}, \citenamefont {Umstadter},\
  and\ \citenamefont {Hulin}}]{Rousse_2004}%
  \BibitemOpen
  \bibfield  {author} {\bibinfo {author} {\bibfnamefont {A.}~\bibnamefont
  {Rousse}}, \bibinfo {author} {\bibfnamefont {K.~T.}\ \bibnamefont {Phuoc}},
  \bibinfo {author} {\bibfnamefont {R.}~\bibnamefont {Shah}}, \bibinfo {author}
  {\bibfnamefont {A.}~\bibnamefont {Pukhov}}, \bibinfo {author} {\bibfnamefont
  {E.}~\bibnamefont {Lefebvre}}, \bibinfo {author} {\bibfnamefont
  {V.}~\bibnamefont {Malka}}, \bibinfo {author} {\bibfnamefont
  {S.}~\bibnamefont {Kiselev}}, \bibinfo {author} {\bibfnamefont
  {F.}~\bibnamefont {Burgy}}, \bibinfo {author} {\bibfnamefont {J.-P.}\
  \bibnamefont {Rousseau}}, \bibinfo {author} {\bibfnamefont {D.}~\bibnamefont
  {Umstadter}}, \ and\ \bibinfo {author} {\bibfnamefont {D.}~\bibnamefont
  {Hulin}},\ }\href@noop {} {\bibfield  {journal} {\bibinfo  {journal} {Phys.
  Rev. Lett.}\ }\textbf {\bibinfo {volume} {93}},\ \bibinfo {pages} {135005}
  (\bibinfo {year} {2004})}\BibitemShut {NoStop}%
\bibitem [{\citenamefont {Cipiccia}\ \emph {et~al.}(2011)\citenamefont
  {Cipiccia}, \citenamefont {Islam}, \citenamefont {Ersfeld}, \citenamefont
  {Shanks}, \citenamefont {Brunetti}, \citenamefont {Vieux}, \citenamefont
  {Yang}, \citenamefont {Issac}, \citenamefont {Wiggins}, \citenamefont
  {Welsh}, \citenamefont {Anania}, \citenamefont {Maneuski}, \citenamefont
  {Montgomery}, \citenamefont {Smith}, \citenamefont {Hoek}, \citenamefont
  {Hamilton}, \citenamefont {Lemos}, \citenamefont {Symes}, \citenamefont
  {Rajeev}, \citenamefont {Shea}, \citenamefont {Dias},\ and\ \citenamefont
  {Jaroszynski}}]{Cipiccia_2011}%
  \BibitemOpen
  \bibfield  {author} {\bibinfo {author} {\bibfnamefont {S.}~\bibnamefont
  {Cipiccia}}, \bibinfo {author} {\bibfnamefont {M.~R.}\ \bibnamefont {Islam}},
  \bibinfo {author} {\bibfnamefont {B.}~\bibnamefont {Ersfeld}}, \bibinfo
  {author} {\bibfnamefont {R.~P.}\ \bibnamefont {Shanks}}, \bibinfo {author}
  {\bibfnamefont {E.}~\bibnamefont {Brunetti}}, \bibinfo {author}
  {\bibfnamefont {G.}~\bibnamefont {Vieux}}, \bibinfo {author} {\bibfnamefont
  {X.}~\bibnamefont {Yang}}, \bibinfo {author} {\bibfnamefont {R.~C.}\
  \bibnamefont {Issac}}, \bibinfo {author} {\bibfnamefont {S.~M.}\ \bibnamefont
  {Wiggins}}, \bibinfo {author} {\bibfnamefont {G.~H.}\ \bibnamefont {Welsh}},
  \bibinfo {author} {\bibfnamefont {M.-P.}\ \bibnamefont {Anania}}, \bibinfo
  {author} {\bibfnamefont {D.}~\bibnamefont {Maneuski}}, \bibinfo {author}
  {\bibfnamefont {R.}~\bibnamefont {Montgomery}}, \bibinfo {author}
  {\bibfnamefont {G.}~\bibnamefont {Smith}}, \bibinfo {author} {\bibfnamefont
  {M.}~\bibnamefont {Hoek}}, \bibinfo {author} {\bibfnamefont {D.~J.}\
  \bibnamefont {Hamilton}}, \bibinfo {author} {\bibfnamefont {N.~R.~C.}\
  \bibnamefont {Lemos}}, \bibinfo {author} {\bibfnamefont {D.}~\bibnamefont
  {Symes}}, \bibinfo {author} {\bibfnamefont {P.~P.}\ \bibnamefont {Rajeev}},
  \bibinfo {author} {\bibfnamefont {V.~O.}\ \bibnamefont {Shea}}, \bibinfo
  {author} {\bibfnamefont {J.~M.}\ \bibnamefont {Dias}}, \ and\ \bibinfo
  {author} {\bibfnamefont {D.~A.}\ \bibnamefont {Jaroszynski}},\ }\href@noop {}
  {\bibfield  {journal} {\bibinfo  {journal} {Nat. Phys.}\ }\textbf {\bibinfo
  {volume} {7}},\ \bibinfo {pages} {867} (\bibinfo {year} {2011})}\BibitemShut
  {NoStop}%
\bibitem [{\citenamefont {Graves}\ \emph {et~al.}(2014)\citenamefont {Graves},
  \citenamefont {Bessuille}, \citenamefont {Brown}, \citenamefont {Carbajo},
  \citenamefont {Dolgashev}, \citenamefont {Hong}, \citenamefont {Ihloff},
  \citenamefont {Khaykovich}, \citenamefont {Lin}, \citenamefont {Murari},
  \citenamefont {Nanni}, \citenamefont {Resta}, \citenamefont {Tantawi},
  \citenamefont {Zapata}, \citenamefont {K\"artner},\ and\ \citenamefont
  {Moncton}}]{Graves_2014}%
  \BibitemOpen
  \bibfield  {author} {\bibinfo {author} {\bibfnamefont {W.~S.}\ \bibnamefont
  {Graves}}, \bibinfo {author} {\bibfnamefont {J.}~\bibnamefont {Bessuille}},
  \bibinfo {author} {\bibfnamefont {P.}~\bibnamefont {Brown}}, \bibinfo
  {author} {\bibfnamefont {S.}~\bibnamefont {Carbajo}}, \bibinfo {author}
  {\bibfnamefont {V.}~\bibnamefont {Dolgashev}}, \bibinfo {author}
  {\bibfnamefont {K.-H.}\ \bibnamefont {Hong}}, \bibinfo {author}
  {\bibfnamefont {E.}~\bibnamefont {Ihloff}}, \bibinfo {author} {\bibfnamefont
  {B.}~\bibnamefont {Khaykovich}}, \bibinfo {author} {\bibfnamefont
  {H.}~\bibnamefont {Lin}}, \bibinfo {author} {\bibfnamefont {K.}~\bibnamefont
  {Murari}}, \bibinfo {author} {\bibfnamefont {E.~A.}\ \bibnamefont {Nanni}},
  \bibinfo {author} {\bibfnamefont {G.}~\bibnamefont {Resta}}, \bibinfo
  {author} {\bibfnamefont {S.}~\bibnamefont {Tantawi}}, \bibinfo {author}
  {\bibfnamefont {L.~E.}\ \bibnamefont {Zapata}}, \bibinfo {author}
  {\bibfnamefont {F.~X.}\ \bibnamefont {K\"artner}}, \ and\ \bibinfo {author}
  {\bibfnamefont {D.~E.}\ \bibnamefont {Moncton}},\ }\href@noop {} {\bibfield
  {journal} {\bibinfo  {journal} {Phys. Rev. ST Accel. Beams}\ }\textbf
  {\bibinfo {volume} {17}},\ \bibinfo {pages} {120701} (\bibinfo {year}
  {2014})}\BibitemShut {NoStop}%
\end{thebibliography}%

\end{document}